\documentclass[10pt,twocolumn,aps,prb,superscriptaddress]{revtex4-1}
\usepackage{graphicx}
\usepackage{dcolumn}
\usepackage{bm}
\usepackage{amsmath,amssymb,latexsym,amsfonts}
\usepackage{amsmath}
\usepackage{yfonts}
\usepackage{graphics,graphicx}
\usepackage{feynmp-auto}

\begin{document}
\title{Decision making by Berry's flux}
\author{Benjamin O. Sung}
\affiliation{Dept. of Physics, Cornell University, Ithaca, NY 14853}
\author{Michael J. Lawler}
\affiliation{Dept. of Physics, Applied Physics and Astronomy, Binghamton University, 
Vestal, NY 13850}
\affiliation{Dept. of Physics, Cornell University, Ithaca, NY 14853}

\date{\today}
\begin{abstract}
Order by disorder is a decision making process for frustrated systems but often leads to simple answers. We study order by disorder in the kagome Kondo model known for its complexity seeking rich decision making capabilities.  At half filling and large Kondo coupling to hopping ratio $J_K/t$, the full manifold of 120$^o$ kagome ground states are degenerate at second order in $t/J_K$. We show this degeneracy lifts at sixth order when a fermion can hop around a hexagon and feel the Berry flux induced by a given spin texture. Using Monte-Carlo we then seek the ground state of this sixth order Hamiltonian and find in a 4x4 unit cell system that a co-planar 2x4 unit cell order is selected over the $\sqrt{3}\times\sqrt{3}$, $q=0$ and cuboc1 states, a result that survives even in the thermodynamic limit. This state is selected for its SU(2) flux properties induced by the spin texture and is analogous to the integer quantum Hall effect. Given the existence of numerous quantum Hall plateaus in other systems, the existence of this 2x4 unit cell state suggests that complex decision making is possible on the manifold of 120$^o$ states and achievable in different kagome-Kondo models.
\end{abstract}
\pacs{Valid PACS appear here}
\maketitle

\section{Introduction}
Order by disorder\cite{Villain1980} is a decision making process: different ground states will be selected from the same degenerate manifold depending on the type of fluctuations it experiences. The ground state with the highest entropy, for example, is selected when the thermal order by disorder effect is active. The ground state with the least zero point energy, on the other hand, is selected when the quantum order by disorder effect is active. These examples are pedagogically discussed in a recent review\cite{Chalker2011}. In effect, the system computes the ground state according to the active order by disorder mechanism. In this way, the order by disorder effect fits into the subject of complexity science\cite{Mitchell2009}.

Yet, in most cases, order by disorder, either thermal, quantum or another mechanism, selects a simple ground state. Naturally, this is expected in the simplest cases such as the J1-J2 model on the square lattice\cite{Chandra1988,Henley1989}. But for kagome antiferromagnets the coplanar $\sqrt{3}\times\sqrt{3}$ state is selected (or prefered if long range order can't be established) both by thermal fluctuations\cite{Chalker:1992p6409} and quantum fluctuations\cite{Chubukov1993a,Henley1995}. This state has a tripled unit cell but is among the simplest in the massively degenerate kagome ground state manifold. Even quantum order by disorder in pyrochlore Heisenberg antiferromagnets selects a colinear state\cite{Henley2006} (though which colinear state is not clear at present\cite{Hizi2009}). Again, this is simpler than a general state in the massively degenerate manifold would suggest. So order by disorder seems to act as a de-complexifying mechanism.

In this light, the discovery of order by disorder in classical Kondo lattice models on highly frustrated lattices\cite{Ghosh2014} is interesting. Complex orders, some of which have multiple wave vectors, are non-coplanar and incommensurate, arise in these models on the square lattice\cite{Ozawa2015}, cubic lattice\cite{Pradhan2009}, triangluar lattice\cite{Ishizuka2012}, between the triangular and kagome lattices\cite{Akagi2013} and kagome lattice\cite{Udagawa2013,Ghosh2014,Barros2014}. Further, the order-by-disorder effect on the kagome lattice model does not select either the $\sqrt{3}\times\sqrt{3}$ order or $q=0$ but possibly\cite{Ghosh2014} the cuboc1 non-coplanar 120$^0$ state\cite{Messio2012}. So it seems possible that order by disorder due to the fermion hopping in these models may give rise to complex selection among a highly degenerate ground state manifold and that it is not de-complexifying. 

Given the potential for complex orders, order-by-disorder in classical Kondo models could also be interesting should it produce an integer quantum Hall effect. This is possible\cite{Ye1999,Ohgushi1999} and indeed has provided much of the motivation for the study of these models\cite{Nagaosa2006}. In 2D an effect is particularly expected at finite temperature should the complex order have a non-zero scalar spin chirality\cite{Martin2008,Bulaevskii2008}. So, if the finite spin chirality\cite{Messio2012,Gong2015} cuboc1 state were the selected state in the kagome case order-by-disorder would provide a mechanism for the stability of a state with an integer quantum Hall effect. 

In this paper, we revisit the order by disorder problem in the kagome Kondo lattice model with classical spins at half filling. This problem is characterized by a small parameter $t/J_K$, a gap to electronic excitations and an SU(2) flux variable ${\bf U}$ felt by the electrons as they hop around in a background classical spin texture. By carrying out a perturbative expansion in $t/J_K$ to sixth order, we show that the selection of a 120$^o$ state is due to the flux felt by an electron as it hops around a hexagon. Using Monte-carlo we then show that the state selected in a $4x4$ unit cell system cluster with periodic boundary conditions has a $2x4$ unit cell. This state turns out to be precisely in between a $\sqrt{3}\times\sqrt{3}$ state and $q=0$ state: it has a spin origami sheet\cite{Shender1993a,Chandra1993} that is fully folded in one direction and perfectly flat in the other. We have verified that it beats the $\sqrt{3}\times\sqrt{3}$, $q=0$ and cuboc1 state in the thermodynamic limit. Further, its SU(2) flux properties are also special: yielding energetic benefits both for hopping around hexagons and on bow-ties (pairs of triangles). Finally, we have computed the electronic band structure and verified that the absence of an integer quantum Hall effect as expected due to the vanishing scalar spin charility. We conclude with an outlook on how these results may generalize to enable selection of other complex ordering patterns within the kagome 120$^o$ states and thereby achieve complex decision making among this manifold of states.

\section{Model}
We begin with the Kondo Hamiltonian given by \begin{equation}\label{eq:kondoham}
H = H_{\text{hop}} + H_{\text{kondo}} = -t\sum_{\langle ij \rangle}c_{i\sigma}^{\dagger}c_{j\sigma} - J_{k}\sum_{i}S_{i}\cdot s_{i}
\end{equation}
Here, the kondo term in the above Hamiltonian gives the coupling between the on-site classical spin vector $S_i$ with the local electron spin $s_i = c^\dagger_i\tau c^{\ }_i$ with coupling constant $J_{k}$. The first term above describes the electron hopping along adjacent sites on the given lattice with amplitude $t$. 

It is well known that in the double exchange model, i.e. the limit $J_{k} \to \infty$, the ferromagnetic state dominates over all other spin states at all but half filling. There, instead of ferromagnetism, a gap in the fermion spectra opens up and antiferromagnetism is found\cite{Ghosh2014}. The leading antiferromagnetic term turns out to be just a nearest neighbor Heisenberg model. 

\begin{figure}[ht]
\includegraphics[width = \columnwidth]{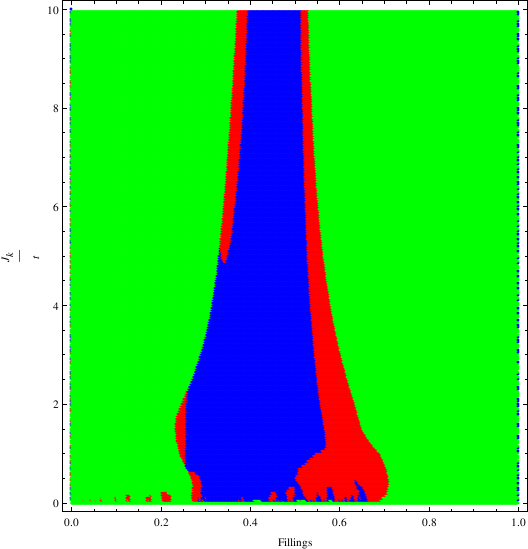}
\caption{Simplified variational phase diagram of kagome Kondo model with classical spins. Green: Ferromagnetic, Blue: $q = 0$, Red: $q = \sqrt{3} \times \sqrt{3}$. Over 1728 sites where there are 24 triangles along the x-axis and 24 triangles along the y-axis. The horizontal axis represents the fraction of electrons occupying a site. The vertical axis is the ratio $\frac{J_{k}}{t}$.}
\label{fig:fig1}
\end{figure}

A quick exploration of the landscape of magnetism is easily achieved using a variational calculation that takes into account just a few possible states: the ferromagnetic, $q = 0$, and $q = \sqrt{3} \times \sqrt{3}$ states. These states are all coplanar. By exact diagonalizing the above hamiltonian on a finite $24 \times 24$ kagome lattice with periodic boundary conditions, we obtain phase diagram of Fig. \ref{fig:fig1}.

In fig.~\ref{fig:fig1}, we have only portrayed the simplest of states of many well-known states on the Kagome lattice with a focus on the general competition between ferromagnetism and antiferromagnetism. The figure suggests that for $\frac{J_{k}}{t} \gg 1$, all states tend to converge to the same energy at half filling in the double exchange model. However, one can slightly weaken this condition, and only consider the approximation $\frac{J_{k}}{t} \gg 1$ and consider which states dominate in this regime. To second order in perturbation theory in $\frac{J_{k}}{t}$, all $120^{o}$ degree states are degenerate\cite{Ghosh2014}. 

In addition to $q=0$ and $\sqrt{3}\times\sqrt{3}$ states, there are an infinite number of other $120^{o}$ spin configurations, both coplanar and non-coplanar, that one may place on an arbitrarily large, finite kagome lattice, and to second order in perturbation theory are all degenerate in energy. In order to extract a true, unique ground state out of the degenerate $120^{o}$ spin state manifold, we must proceed to higher orders in perturbation theory. 

\section{Variational Results} 
In this paper, we will show that the degeneracy is lifted exactly at sixth order and exhibit a newly found state with a $4\times 2 $ unit cell via a monte carlo simulation that beats the above well-known $120^{o}$ states. Here, we can begin to understand this result with numerical evidence that the degeneracy is lifted at sixth order.

Our numerical argument follows by presenting the energy data for three well-known $120^{o}$ states. The energy data for these spin configurations were obtained by exact diagonalizing the hamiltonian given in equation~\ref{eq:kondoham} and summing the lower half of the eigenvalue spectrum, corresponding to half filling. On an $24 \times 24$ kagome lattice, the first few energy values in terms of the coupling constant $J_{k}$ are as in the table below:
\\
\begin{center}
\begin{tabular}{|c|c|c|c|}
\hline
$J_{k}$ & q = 0 & $q = \sqrt{3} \times \sqrt{3}$ & cuboc1\\
\hline
1 & -3388.466 & -3409.835 & -3387.137\\
\hline
2 & -4530.299 & -4535.492 & -4535.302 \\
\hline
3 & -5967.679 & -5968.139 & -5968.586\\
\hline
4 & -7522.879 & -7522.914 & -7523.120 \\
\hline
5 & -9138.454 & -9138.446 & -9138.537\\
\hline
\end{tabular}
\end{center}
We easily see that at small $J_{k}$, the energy data for the three $120^{o}$ states differ slightly, due to non-trivial subleading terms in higher-order perturbation theory. However, as we increase $J_{k}$, terms of order $n$ in perturbation theory are suppressed by the factor $\frac{1}{J_{k}^{n-1}}$, and hence the energies begin to converge. By $J_k/t = 5$, the cuboc1 state is winning but only in the 5th significant figure. 

\begin{figure}[t]
\includegraphics[width=\linewidth]{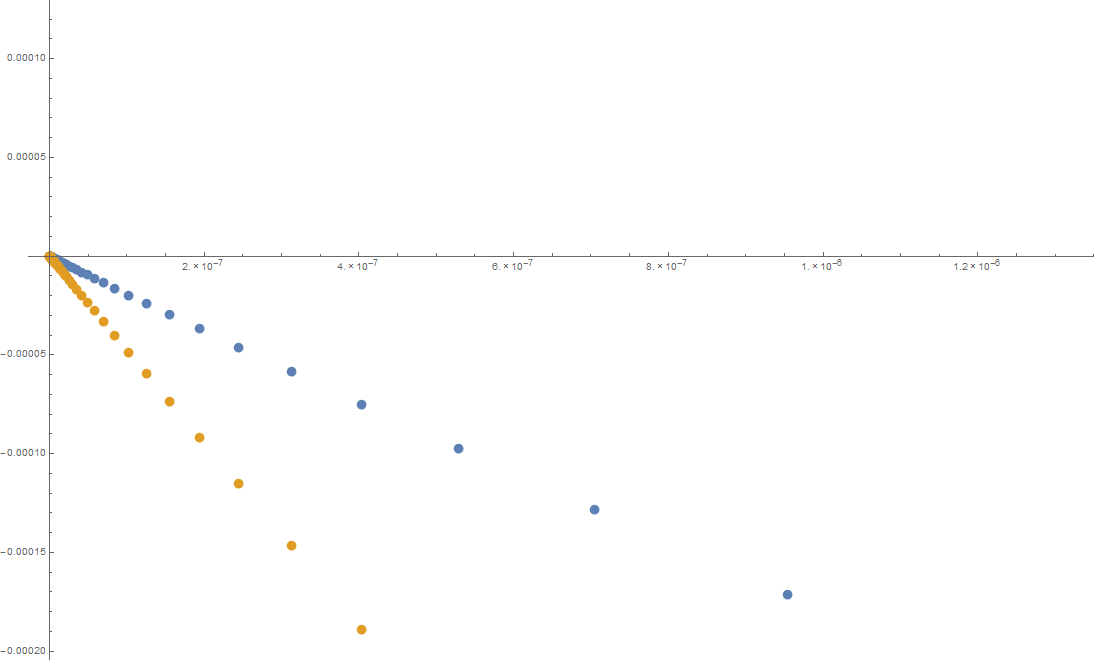} 
\caption{Graph of $J_{k}$ vs Energy. The energy values from $J_{k} = 1$ to 100 are plotted. The blue dots correspond to the difference between the states $q=0$ and $q=\sqrt{3} \times \sqrt{3}$ and the yellow dots correspond to the difference between the states cuboc1 and $q = \sqrt{3} \times \sqrt{3}$ states. The x-axis is $\frac{1}{J_{k}^{5}}$ as $J_{k}$ goes from 1 to 100, and the y-axis are the energy values. The linearity of the two plots shows that the degeneracy breaks first at sixth order in perturbation theory. Energy calculations were run on a 24x24 kagome lattice.}
\label{fig:fig2}
\end{figure}

We now extend this numerical evidence for degeneracy splitting among the $120^{o}$ states at sixth order. By acquiring energy data as in the above for the spectrum $J_{k} = 1$ to $100$, and taking the differences in energies, we obtain figure~\ref{fig:fig2} and clear evidence that the degeneracy is lifted at sixth order.

Having shown numerically that degeneracy breaks at sixth order, one may calculate the contributions from each order in perturbation theory via numerical fitting. However, due to noise, it is nearly impossible to calculate the numerical coefficients to higher orders above 4. Further, even though we expect that the coefficients at third and fifth order vanish, the fifth order term as calculated numerically suggests a probable non-trivial coefficient. By carrying out an asymptotic fitting of the energy data we obtain the table: 
\\
\begin{center}
\begin{tabular}{|c|c|c|}
\hline
	Order of $\frac{1}{J_{k}}$ &  Numerical Coefficients\\
\hline
	1 & -2591.73\\
\hline
	2 & 45.4979\\
\hline
	3 & 2751.6 \\
\hline
	4 & -233.15\\
\hline
\end{tabular}
\end{center}
This numerical data would not be sufficient to determine the exact formula for any contributions from higher orders in perturbation theory. But it does provide us with a check on our analytic calculation below. 

\section{Feynman Diagram Approach}
\subsection{Roadmap of Feynman Diagram Calculation}
Here, we provide a roadmap for our feynman diagram calculation. We proceed through the canonical method, writing down the path integral $Z$ using our free Hamiltonian eqn~\ref{eq:kondoham} without the hopping term. As usual, completing the square yields the propagator for the free theory. Adding the interaction (hopping term) and taking functional derivatives yields the full propagator for our theory. By taylor expanding in the perturbation $H_{\text{hop}}$, we obtain the interaction $U_{ij}$, which may be readily computed via unitary diagonalization as a $2 \times 2$ matrix describing the hopping between nearest neighbor sites. We then  calculate the energy corrections via the formula eqn~\ref{eq:feynenergy} which was calculated using the linked cluster theorem. We further highlight the procedure for the first two orders in perturbation theory and show that they agree with results from the usual quantum mechanical procedure.
\subsection{Derivation of Feynman Rule for Propagator}
We begin our diagrammatic procedure by deriving the feynman rules for our theory using the canonical approach. Proceeding via path integral quantization with grassman variables, we take the kondo part of our hamiltonian $$H_{\text{kondo}} = -J_{k}\sum_{\langle ij \rangle}S_{i}\cdot s_{i}$$ and obtain 
\begin{equation}\begin{split}
Z &= \int DcD\bar{c}\>exp(i\int dt[\sum_{i}i\bar{c}_{i\sigma}\partial_{t}c_{i\sigma} - H])\\
&= \int DcD\bar{c}\>exp(i\int dt[\sum_{i}i\bar{c}_{i\sigma}\partial_{t}c_{i\sigma}\\ 
&+ J_{k}\sum_{i}S_{i}\cdot \bar{c}_{i\sigma}\tau_{\sigma \sigma'}c_{i\sigma'} + \sum_{i}(\bar{\eta}_{i\sigma}c_{i\sigma} + \bar{c}_{i\sigma}\eta_{i\sigma}))
\end{split}\end{equation}
To calculate the propagator for our free theory, we Fourier transform to momentum space and complete the square, obtaining the green's function given below 
\begin{equation}
G(k,\omega) = \frac{1}{i\omega + J_{k}\tau_{3} - i\epsilon}
\end{equation}
Calculating the propagator for the interacting theory follows as follows: we consider the hopping term $$H_{\text{hop}} = -t\sum_{\langle ij \rangle}tc_{i\sigma}^{\dagger}c_{j\sigma}$$ Taking functional derivatives of the path integral with respect to the grassman valued sources $\bar{\eta}_{i\sigma}$ and $\eta_{i\sigma}$ allows us to make the replacement 
\begin{equation}
H_{i} \to \sum_{\langle ij \rangle}\frac{\delta}{\delta \eta_{i\sigma}}U_{ij}^{\sigma \sigma'}\frac{\delta}{\delta \bar{\eta}_{j\sigma'}}
\end{equation}
where $U_{ij}^{\sigma \sigma'}$ describes the hopping between nearest neighbor sites. This gives us the full green's function  
\begin{equation}
G^{\sigma_{1}\sigma_{2}}_{j_{1}i_{2}} = \frac{\delta^{\sigma_{1}\sigma_{2}}_{j_{1}i_{2}}}{\omega - J_{k}\tau^{3} + i\epsilon sgn(\omega)}
\end{equation}
where the upper indices denote spin indices, and the lower indices denote points in position space.

\subsection{Derivation of Feynman Rule for Interaction Vertex}
Now that we have calculated the propagator for our theory, we simply need to determine the feynman rule for our interaction vertex. To do this, note that we must calculate the amplitude for hopping between two sites given by some unitary $2 \times 2$ matrix $U_{ij}$, corresponding to up and down spin states. We begin with equation~\ref{eq:kondoham} in the matrix representation
\begin{equation}
H=\begin{pmatrix}
-J_{k}S^{z} & -J_{k}S^{x}-iJ_{k}S^{y}\\
	-J_{k}S^{x}+iJ_{k}S^{y} & J_{k}S^{z}\\
\end{pmatrix}
\end{equation}
Diagonalizing as usual, we obtain the unitary matrices
\begin{equation}
U = \begin{pmatrix}
	\frac{-S^{x}-iS^{y}}{\sqrt{2+2S^{z}}} & \frac{S^{x}+iS^{y}}{\sqrt{2-2S^{z}}}\\
	\frac{S^{z}+1}{\sqrt{2+2S^{z}}} & \frac{-S^{z}+1}{\sqrt{2-2S^{z}}}\\
\end{pmatrix}, 
	U^{\dagger} = \begin{pmatrix}
	\frac{-S^{x}+iS^{y}}{\sqrt{2+2S^{z}}} & \frac{S^{z}+1}{\sqrt{2+2S^{z}}}\\
	\frac{S^{x}-iS^{y}}{\sqrt{2-2S^{z}}} & \frac{-S^{z}+1}{\sqrt{2-2S^{z}}}\\
\end{pmatrix}
\end{equation}
Expressing the hopping $H_{hop}$ in terms of unitary matrices, we obtain
\begin{equation}
H_{1} = -t\sum_{<ij>}(U^{i\dagger}_{\sigma\sigma'}U^{j}_{\sigma\sigma''})c^{\dagger}_{i\sigma'}c_{j\sigma''} + h.c.
\end{equation}
Finally, writing out the product of the unitary matrices explicitly, for hopping between two sites with classical spin vectors $S_{i}$ and $S_{j}$, we obtain 
\begin{equation}\begin{split}
U_{11} & = 
\frac{1}{\sqrt{2+2S^{z^{i}}}\sqrt{2+2S^{z^{j}}}}(S^{x^{i}}S^{x^{j}}-iS^{y^{i}}S^{x^{j}} + i S^{x^{i}}S^{y^{j}}\\ 
&+ S^{y^{i}}S^{y^{j}} + S^{z^{i}}S^{z^{j}}+S^{z^{i}}+S^{z^{j}}+1) \\
U_{12} & = \frac{1}{\sqrt{2+2S^{z^{i}}}\sqrt{2-2S^{z^{j}}}}(-S^{x^{i}}S^{x^{j}}+iS^{y^{i}}S^{x^{j}}-iS^{y^{j}}S^{x^{i}}\\ 
&- S^{y^{i}}S^{y^{j}} - S^{z^{i}}S^{z^{j}} + S^{z^{i}} - S^{z^{j}} + 1)\\
U_{21} & = \frac{1}{\sqrt{2-2S^{z^{i}}}\sqrt{2+2S^{z^{j}}}}(-S^{x^{i}}S^{x^{j}} + iS^{y^{i}}S^{x^{j}} - iS^{x^{i}}S^{y^{j}} \\
&- S^{y^{i}}S^{y^{j}}- S^{z^{i}}S^{z^{j}} + S^{z^{j}} - S^{z^{i}} + 1)\\
U_{22} & = \frac{1}{\sqrt{2-2S^{z^{i}}}\sqrt{2-2S^{z^{j}}}}(S^{x^{i}}S^{x^{j}} - iS^{y^{i}}S^{x^{j}} + iS^{x^{i}}S^{y^{j}} \\
&+ S^{y^{i}}S^{y^{j}} +S^{z^{i}}S^{z^{j}} - S^{z^{i}} - S^{z^{j}} + 1)
\end{split}\end{equation}

\subsection{Energy Calculation}
In full, the above calculation gives us the following feynman rules:
\begin{center}
$G_{ij}^{\sigma \sigma'}$
\end{center}
\begin{center}
\begin{fmffile}{Propagator}
\begin{fmfgraph*}(120,10)
\fmfleft{i}
\fmfright{o}
\fmf{plain}{i,v,o}
\fmflabel{$i,\sigma$}{i}
\fmf{dot}{i}
\fmflabel{$j,\sigma'$}{o}
\fmf{dot}{o}
\end{fmfgraph*}
\end{fmffile}
\end{center}
 
\vspace{4mm}

\begin{center}
$U_{ij}^{\sigma \sigma'}$
\end{center}
\begin{center}
\begin{fmffile}{Interaction}
\begin{fmfgraph*}(120,10)
\fmfleft{i}
\fmfright{o}
\fmf{photon}{i,v,o}
\fmflabel{$i,\sigma$}{i}
\fmf{dot}{i}
\fmflabel{$j,\sigma'$}{o}
\fmf{dot}{o}
\end{fmfgraph*}
\end{fmffile}
\end{center}

To calculate an $n$th order process, the Taylor expansion tells us to connect $n$ G's and $n$ U's in an alternating order. Intuitively, one may think of the propagator as the virtual electron hopping between nearest neighbors, and the interaction $U$ as imposing the constraint that the path traversed by the virtual electron must be connected.  

Before moving to sixth order, we give a sample calculation to first and second order. We begin with first order, to check that our path integral does indeed give zero for only one hopping.

\vspace{7mm}

\begin{center}
\begin{fmffile}{1stOrder}
\begin{fmfgraph*}(120,40)
\fmfleft{i}
\fmfright{o}
\fmf{plain,tension=5}{i,v1}
\fmf{plain,tension=5}{v2,o}
\fmf{plain,left,tension=0.4}{v1,v2}
\fmf{photon}{v1,v2}
\fmfdot{v1,v2}
\fmflabel{$i,\sigma$}{i}
\fmflabel{$j,\sigma'$}{o}
\end{fmfgraph*}
\end{fmffile}
\end{center}
The expression to calculate the energy is given by 
\begin{equation}\begin{split}\label{eq:feynenergy}
E &= -\frac{i}{T}ln(Z_{0}) - \frac{i}{T}ln(\langle exp(-i\int dt H_{I})\rangle)\\
&= E_{G}^{(0)} - \frac{i}{T}ln(\langle exp(-\frac{i}{T}\int dt H_{I}\rangle_{0}\\
\end{split}\end{equation}
Hence, proceeding with the calculation for first order processes, we obtain 
\begin{equation}\begin{split}
E' & = \int d\tau -t \sum_{ij} \langle \overline{c_{i\sigma}(t)}U_{ij}^{\sigma \sigma'}c_{j\sigma'}(t)\rangle \\
&= t Tr\int_{t_{1}}^{t_{2}} d\tau \sum_{ij}G_{ij}(t - t)U_{ij}\\
&= tT\sum_{ij}\int \frac{d\omega}{2\pi}G_{ij}(\omega)U_{ij}
\end{split}\end{equation}
Now, noting that the green's function gives the delta function $G_{ij} \propto \delta_{ij}$ and that $U_{ij}$ vanishes identically for $i = j$, we see that the correction vanishes. 

We now consider the integral to second order. In this case, we have the diagram

\vspace{7mm}

\begin{center}
\begin{fmffile}{2ndOrder}
\begin{fmfgraph*}(120,40)
\fmfleft{i1,i2}
\fmfright{o1,o2}
\fmf{plain}{i1,v1}
\fmf{plain}{v2,o1}
\fmf{plain}{i2,v3}
\fmf{plain}{v4,o2}
\fmf{photon}{v1,v2}
\fmf{photon}{v3,v4}
\fmfdot{v1,v2}
\fmfdot{v3,v4}
\fmflabel{$i,\sigma_{i}$}{i1}
\fmflabel{$j,\sigma_{j}$}{o1}
\fmflabel{$k,\sigma_{k}$}{i2}
\fmflabel{$l,\sigma_{l}$}{o2}
\end{fmfgraph*}
\end{fmffile}
\end{center}

\vspace{5mm}

Here, we have two vertical lines corresponding to two vertices on the kagome lattice. Although not indicated, we choose the convention with the direction of propagation to the right, hence, we may only connect, using a propagator, a vertex at the left with a vertex at the right. For instance, we may connect $(k,\sigma_{k})$ with $(j,\sigma_{j})$, but not $(k,\sigma_{k})$ with $(i,\sigma_{i})$. Note that there are only two distinct ways to connect using propagators as indicated below:

\vspace{7mm}

\begin{center}
\begin{fmffile}{2ndOrdercon1}
\begin{fmfgraph*}(120,40)
\fmfleft{i1,i2}
\fmfright{o1,o2}
\fmf{plain}{i1,v1}
\fmf{plain}{v2,o1}
\fmf{plain}{i2,v3}
\fmf{plain}{v4,o2}
\fmf{photon}{v1,v2}
\fmf{photon}{v3,v4}
\fmfdot{v1,v2}
\fmfdot{v3,v4}
\fmflabel{$i,\sigma_{i}$}{i1}
\fmflabel{$j,\sigma_{j}$}{o1}
\fmflabel{$k,\sigma_{k}$}{i2}
\fmflabel{$l,\sigma_{l}$}{o2}
\fmf{plain}{i1,o2}
\fmf{plain}{i2,o1}
\end{fmfgraph*}
\end{fmffile}
\end{center}

\vspace{18mm}

\begin{center}
\begin{fmffile}{2ndOrdercon2}
\begin{fmfgraph*}(120,40)
\fmfleft{i1,i2}
\fmfright{o1,o2}
\fmf{plain}{i1,v1}
\fmf{plain}{v2,o1}
\fmf{plain}{i2,v3}
\fmf{plain}{v4,o2}
\fmf{photon}{v1,v2}
\fmf{photon}{v3,v4}
\fmfdot{v1,v2}
\fmfdot{v3,v4}
\fmflabel{$i,\sigma_{i}$}{i1}
\fmflabel{$j,\sigma_{j}$}{o1}
\fmflabel{$k,\sigma_{k}$}{i2}
\fmflabel{$l,\sigma_{l}$}{o2}
\fmf{plain,right,tension=0.05}{i1,o1}
\fmf{plain,left,tension=0.05}{i2,o2}
\end{fmfgraph*}
\end{fmffile}
\end{center}

\vspace{18mm}
\noindent
As shown, the second diagram is disconnected, and hence it does not contribute to the energy correction at second order. Evaluating the first diagram as usual, one can verify that we obtain the correction
\begin{equation}\begin{split}\label{eq:secondorder}
E' &= Tr(\int \frac{d\omega}{2\pi}\sum_{ijkl}G_{li}U_{ij}G_{jk}U_{kl})\\
&= - \frac{3}{8} J_{k}^{2}
\end{split}\end{equation}
where the spin indices have been suppressed where the second line has been evaluated for the case of $120^{o}$ states. This answer agrees with our calculation using the unitary matrices and the usual quantum mechanical perturbative energy formula. 

We may carry this procedure out to sixth order, with which we will obtain the expression 
\begin{equation}\begin{split}\label{eq:sixthorder}
E'^{6} = Tr(\int \frac{d\omega}{2\pi}\sum GUGUGUGUGUGU)
\end{split}\end{equation}
where spatial and and spin indices have been suppressed. We remark that there is no need to connect all the possible lines in a single six order feynman diagram since the energy sums over all possible spatial indices. Hence, with these feynman rules, we may derive perturbative energy corrections to any order with a single feynman diagram. 

\subsection{Numerical Calculations}
Clearly, it is necessary to numerically evaluate the analytical expression obtained from feynman diagrammatic techniques for energy corrections of order $n > 2$. To do this, we separate equation~\ref{eq:sixthorder} into a linear combination of Pauli matrices as in the below.

We define the variables 
\begin{equation}\begin{split}
&\mu = \frac{1}{2}(\frac{1}{\omega - J_{k} + i\epsilon} + \frac{1}{\omega + J_{k} - i\epsilon})\\ 
&\nu = \frac{1}{2}(\frac{1}{\omega - J_{k} + i\epsilon} - \frac{1}{\omega + J_{k} - i\epsilon})
\end{split}\end{equation}
and 
\begin{equation}\begin{split}
\alpha &= \frac{1}{2}(\frac{1}{\sqrt{2+2S_{x_{3}}}\sqrt{2+2S_{y_{3}}}}(S_{x} \cdot S_{y} - iS_{x_{2}}S_{y_{1}}\\ 
&+ iS_{x_{1}}S_{y_{2}} + S_{x_{3}} + S_{y_{3}} + 1) \\
&+ \frac{1}{\sqrt{2-2S_{x_{3}}}\sqrt{2-2S_{y_{3}}}}(S_{x} \cdot S_{y} - iS_{x_{2}}S_{y_{1}}\\ 
&+ iS_{x_{1}}S_{y_{2}} - S_{x_{3}} - S_{y_{3}} + 1))\\
\delta &= \frac{1}{2}(\frac{1}{\sqrt{2+2S_{x_{3}}}\sqrt{2+2S_{y_{3}}}}(S_{x} \cdot S_{y} - iS_{x_{2}}S_{y_{1}}\\ 
&+ iS_{x_{1}}S_{y_{2}} + S_{x_{3}} + S_{y_{3}} + 1) \\
&- \frac{1}{\sqrt{2-2S_{x_{3}}}\sqrt{2-2S_{y_{3}}}}(S_{x} \cdot S_{y} - iS_{x_{2}}S_{y_{1}}\\ 
&+ iS_{x_{1}}S_{y_{2}} - S_{x_{3}} - S_{y_{3}} + 1))\\
\beta &= \frac{1}{2}(\frac{1}{\sqrt{2+2S_{x_{3}}}\sqrt{2-2S_{y_{3}}}}(-S_{x} \cdot S_{y} +iS_{x_{2}}S_{y_{1}} \\
&- iS_{y_{2}}S_{x_{1}} + S_{x_{3}} - S_{y_{3}} + 1) \\
&+ \frac{1}{\sqrt{2-2S_{x_{3}}}\sqrt{2+2S_{y_{3}}}}(-S_{x}\cdot S_{y} +iS_{x_{2}}S_{y_{1}}\\
&-iS_{y_{2}}S_{x_{1}} - S_{x_{3}}+S_{y_{3}}+1))\\
\gamma &= \frac{1}{2}(\frac{1}{\sqrt{2+2S_{x_{3}}}\sqrt{2-2S_{y_{3}}}}(-S_{x} \cdot S_{y} +iS_{x_{2}}S_{y_{1}}\\ 
&- iS_{y_{2}}S_{x_{1}} + S_{x_{3}} - S_{y_{3}} + 1) \\
&- \frac{1}{\sqrt{2-2S_{x_{3}}}\sqrt{2+2S_{y_{3}}}}(-S_{x}\cdot S_{y} +iS_{x_{2}}S_{y_{1}}\\
&-iS_{y_{2}}S_{x_{1}} - S_{x_{3}}+S_{y_{3}}+1))
\end{split}\end{equation}
and rewrite the green's function and unitary matrices as
\begin{equation}\begin{split}
\frac{1}{\omega - J_{k}\tau^{3}_{\sigma 2 \sigma 2}} &= \mu \tau^{0} + \mu \tau^{3}\\
U^{\sigma \sigma'}_{ij} &= \alpha \tau^{0} + \beta \tau^{1} + \delta \tau^{2} + \gamma \tau^{3}\\
\end{split}\end{equation}
Consequently, we may rewrite equation~\ref{eq:sixthorder} by substituting in these linear combinations, and evaluate it numerically in mathematica. We find to third order and fifth order that the energy correction vanishes as expected from the numerical results. 

Intuitively, we may think of the numerical result that degeneracy is first lifted at sixth order as follows. Second order in perturbation theory selects out a degenerate manifold of $120^{o}$ states. Since third order and fifth order vanishes, the fourth order term is the only possible non-trivial contribution. However, if we think of the electron as hopping around paths with the condition that it begins and ends on the same vertex, it is easy to see that there are no trivial loops that could be achieved via four hoppings. This is only viable at sixth order, in which the electron may hop across bowtie loops and hexagon loops. In particular, we verify numerically using the above procedure that fourth order terms are identical for the well-known $120^{o}$ states on the kagome lattice. We attribute the non-trivial sixth order contribution as resulting from the Berry's phase. 

We further proceeded via a simulation on a $4 \times 4$ kagome lattice, on which we evaluated the above expression numerically and verified that the cuboc1 state dominates over the $q = 0$ state at sixth order as expected. Having derived this expression, this motivates us to carry out a Monte-Carlo simulation to find a global minimum of this theory. In the next section, we detail the results that we obtained using a non-linear local optimization method.

\section{Numerical Nonlinear Local Optimization}
Using the expression derived above, we may numerically evaluate the sixth order contributions from any spin configuration on the kagome lattice. To do this, we create an ensemble of $1000$ random spin configurations on the $4 \times 4$ kagome lattice. We chose the system size of a $4 \times 4$ kagome lattice since we want a small size system to simplify calculations. In particular, this is the smallest size system on which we have no non-trivial closed loops that do not occur on the lattice in the infinite size limit. For example, on a $3 \times 3$ kagome lattice, we may have a non-trivial loop due to the periodic boundary conditions that would not be a loop on the infinite size lattice. We then proceed to numerically minimize each random spin configuration by imposing the $120^{o}$ condition, i.e., neighboring on-site spin vectors must have an inner product of $-\frac{1}{2}$ using the NMinimize method in Mathematica. We then calculate the sixth order contribution of each of these $120^{o}$ states. Upon doing this, we encountered a new state, with energy lower than any other well-known state, which we will detail below. 

\begin{figure}[ht]
\centering
\includegraphics[width=\linewidth]{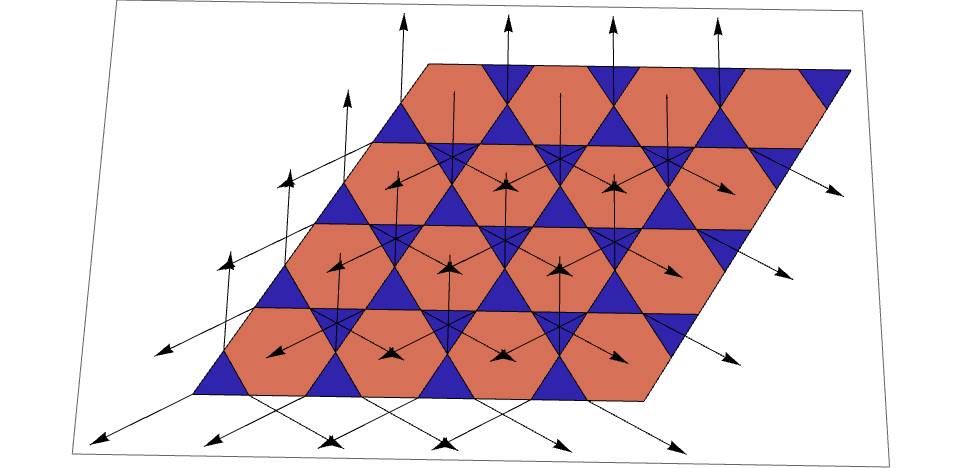}
\includegraphics[width=\linewidth]{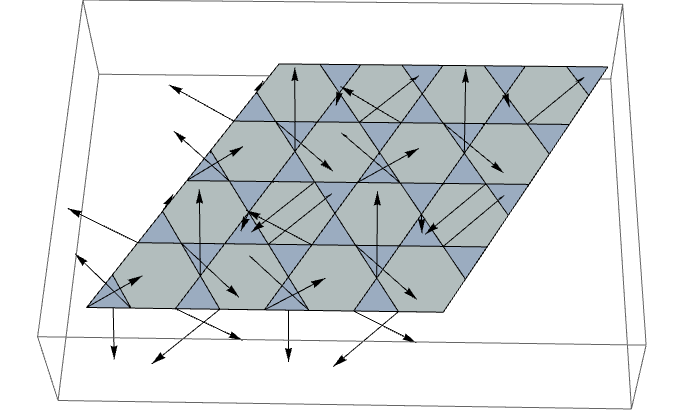}
\includegraphics[width=\linewidth]{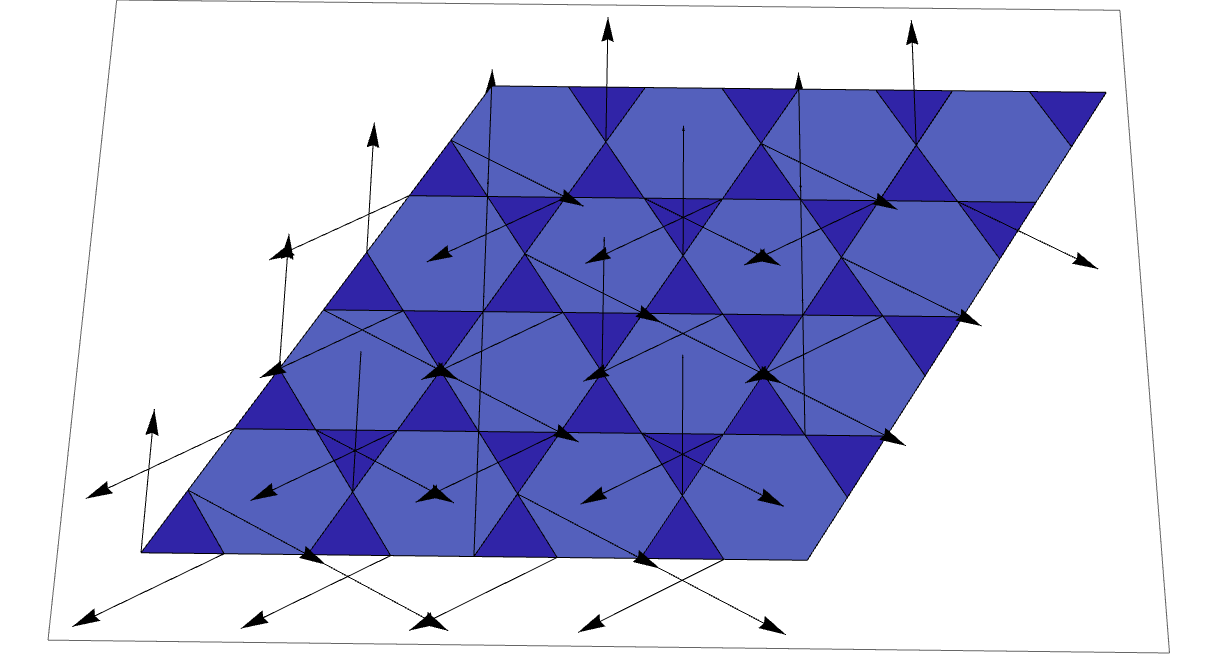}
\caption{Spin pattern plots and their associated hexagon fluxs that contribute at sixth order. The colors correspond to the values of the fluxes through the loops, i.e. "hotter" colors are greater and "colder" colors are smaller. Top sub-figure is the $q=0$ state, middle sug-figure is the cuboc1 state and the bottom sub-figure is the "snake" state.}
\label{fig:globfig}
\end{figure}

By calculating the sixth order contributions from non-trivial loops on each of the states on the $4 \times 4$ kagome lattice, we obtain figure~\ref{fig:globfig} which portrays the relative contributions.

As shown, the sixth order contribution from the ``snake state'' dominates over both the $q = 0$ and $cuboc1$ states. We now proceed to discuss some properties of the newly found state.

\begin{figure}[ht]
\includegraphics[height=5cm,width=\linewidth]{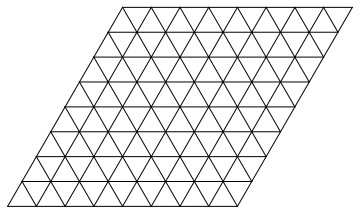}
\includegraphics[height=5cm,width=5.7735cm]{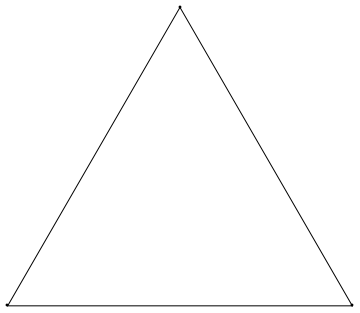}
\includegraphics[height=5cm,width=\linewidth]{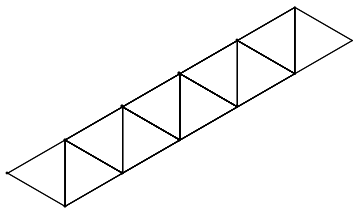}
\caption{Spin origami 
plots\cite{Shender1993a,Chandra1993} that map a 120$^o$ state to a folded sheet of paper. Top state is the $q=0$ pattern that maps to a flat sheet of paper. Middle is the $\sqrt{3}\times\sqrt{3}$ state that maps to a single triangle (completly folded sheet of paper). Bottom is the snake state that maps to a strip that is flat in one direction and completely folded in the other. Each of these patterns corresponds to an $8 \times 8$ kagome lattice with open boundary conditions. As indicated, the ``snake'' spin state exhibits a spin origami pattern that appears to lie between the $q = 0$ and $q = \sqrt{3} \times \sqrt{3}$ spin origami plots.
}
\label{fig:globfig1}
\end{figure}

\begin{figure}[htpb]
\centering
\includegraphics[width=\linewidth]{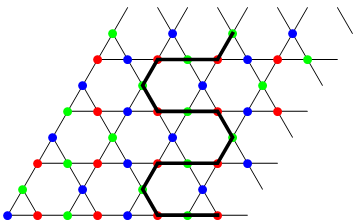} \caption{The newly found state is called the "snake" state due to the above behavior of alternating colors. It is a coplanar state with rgb colors corresponding to the usual spins on the $q = 0$ and $q = \sqrt{3} \times \sqrt{3}$ states.}
\label{fig:snakesplot}
\end{figure}

\begin{figure}[htpb]
\centering
\includegraphics[width=\linewidth]{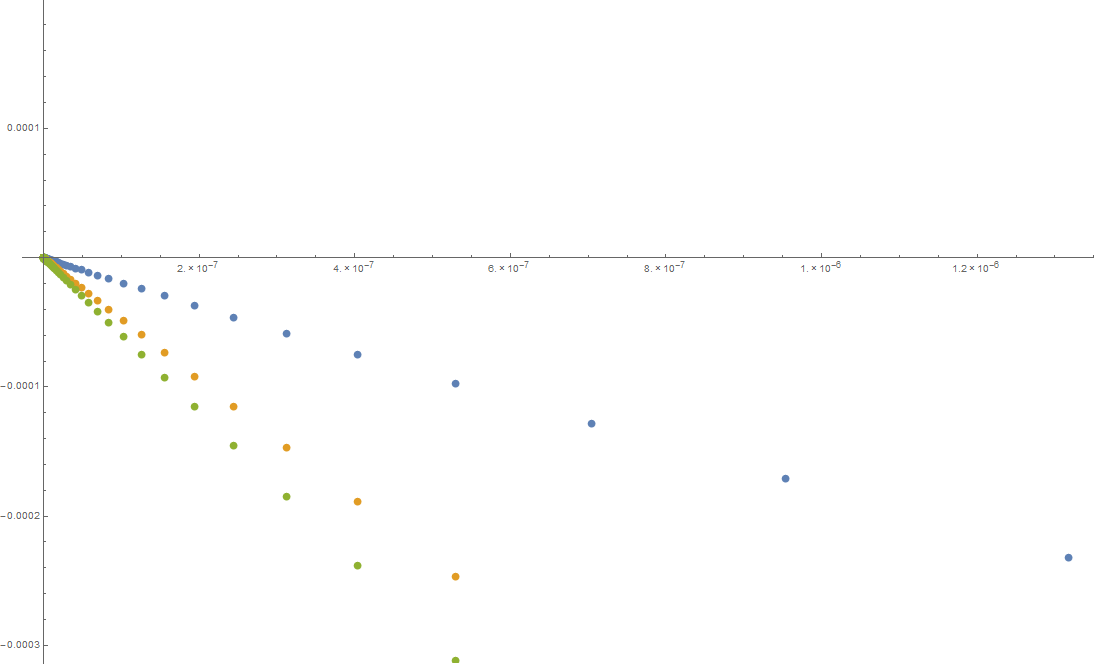} \caption{The additional green line denotes the difference in energies for $J_{k} = 1$ to $100$ between the "snake" and $q = \sqrt{3} \times \sqrt{3}$ states. This clearly shows that to sixth order, degeneracy among the $120^{o}$ states are broken, and our "snake" state is the winner. Energy calculations were run on a 24x24 kagome lattice.}
\label{fig:snakesixthorder}
\end{figure}

Figure~\ref{fig:snakesplot} explains our name for this new found state. The traversal of "ABABAB..." movements take on a "snake-like"  shape on the kagome lattice. Further evidence that the "snake" state is indeed the minimum of all our present states is given by figure~\ref{fig:snakesixthorder} (cf. figure~\ref{fig:fig2}). As clearly shown on the same figure, the "snake" state indeed exhibits lower energies than the $q = 0$, $q = \sqrt{3} \times \sqrt{3}$, and $cuboc1$ states as calculated on the $24 \times 24$ kagome lattice. The reader should note that this verification is completely independent of our perturbation theory. This was calculated using only the Hamiltonian for the system and inputting the relevant data for the classical spin vectors for the ``snake" state.

We now make one further remark regarding the spin plots displayed in figure~\ref{fig:globfig}. For coplanar $120^{o}$ states on the kagome lattice, it can be shown that only hexagonal fluxes contribute to the breaking of degeneracies since the bowtie loops all contribute the same energy. As detailed in the table below, the contributions of hexagonal fluxes is completely consistent with our observations that at sixth order, we have the ordering of states: $\text{snake} < \text{cuboc1} < q = 0 < q = \sqrt{3} \times \sqrt{3}$.

\begin{center}
\begin{tabular}{|c|c|c|c|}
\hline
	Classical State & Bowtie 1 & Bowtie 2 & Hexagon \\
\hline
	$q=\sqrt{3} \times \sqrt{3}$ & 0.158203 & -0.251953 & 0.333984\\
\hline
	$q=0$ & 0.158203 & -0.251953 & 0.158203\\
\hline
	Snake & 0.158203 & -0.251953 & -0.193359\\
\hline
\end{tabular}
\end{center}

\section{Electronic band structure of snake state}
\begin{figure}
\includegraphics[width=\columnwidth]{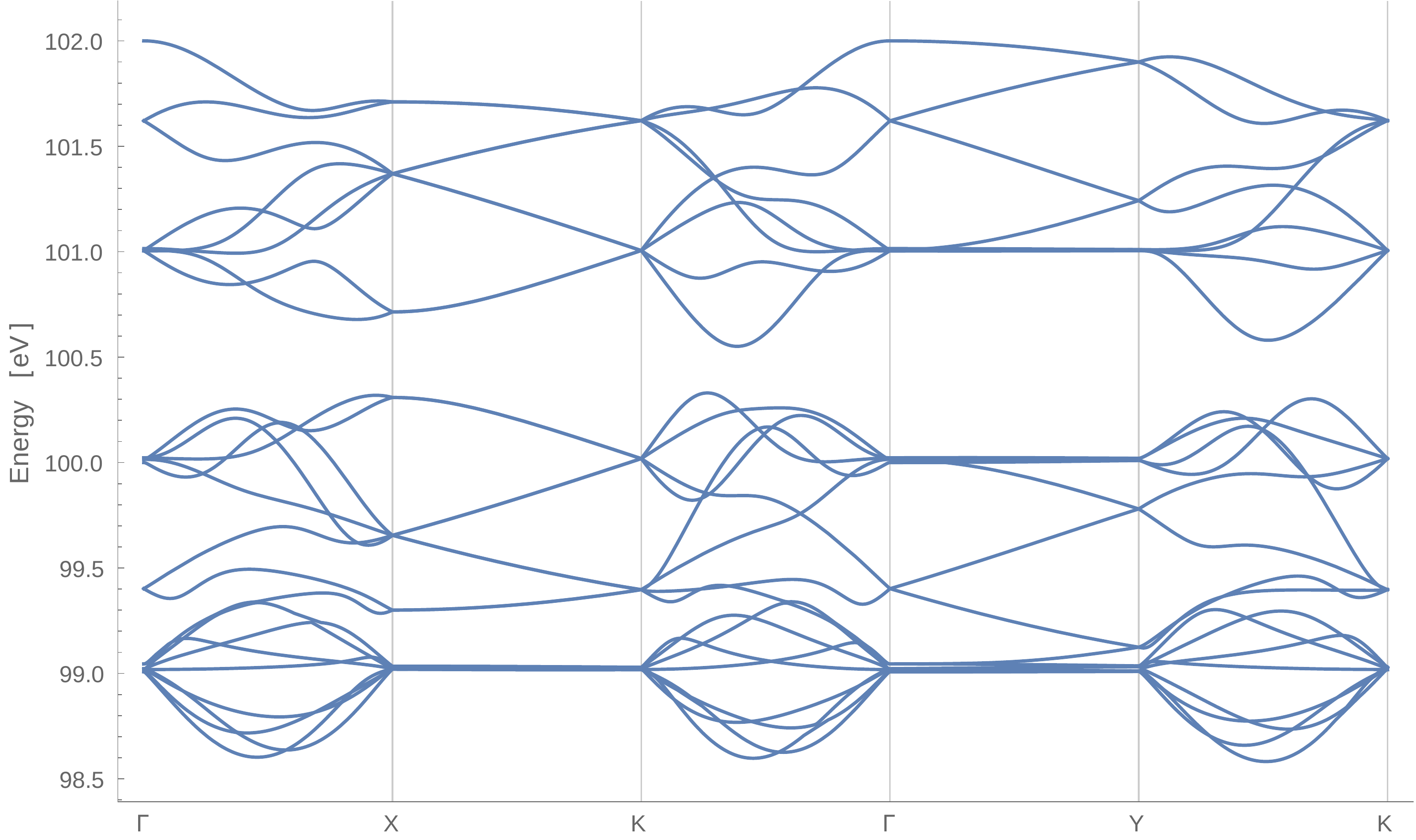}
\caption{Electronic band structure for the electrons hopping in the background of magnetic ordering of the snake state. Here the Brillouin zone is rectangular with sides at the X and Y points and corners at the K point.}
\label{fig:bands}
\end{figure}

Finally, we have computed the electronic band structure of the snake state and its associated Chern number. The band structure for the filled bands is shown in \ref{fig:bands}. We have further computed the Chern number following Ref. \onlinecite{Fukui2005}. We find both for the bottom group of 16 bands and the top group of 8 bands (together making the 24 filled bands out of 48 bands present in the snake state) that the Chern number is trivial. So order by disorder, at least as determined within our 4x4 unit cell calculation, is not selecting a state that could support an integer quantum Hall effect. 

\section{Conclusion}
In this paper, we explored the problem of state selection at half filling of the kagome Kondo model with classical spins. We were motivated to derive an analytical expression for higher orders in perturbation theory in order to better understand the degenerate $120^{o}$ state manifold and achieved this to sixth order.

Proceeding via a numerical non-linear local optimization algorithm, we find, out of an ensemble of 1000 random $120^{o}$ states, a state that beats all the other well-known state as readily verified numerically via fig~\ref{fig:snakesixthorder}. In particular, combining the $120^{o}$ state minimization along with the sixth order expression in the algorithm, we find that many of the runs readily converge to this state. Further work in this direction would include working with a larger ensemble with a more powerful machine in an attempt to find an even better ground state. In particular, we would like to fully understand the contribution of the fluxes to the sixth order correction. While we are drawn to the conclusion that the hexagon fluxes are responsible for the relative correction among the coplanar $120^{o}$ states, we are not completely sure as to how the bowtie and hexagon fluxes contribute to the non-coplanar $120^{o}$ states. 

It is remarkable that the snake state we find has a unit cell with 24 spins (2x4 unit cells) but was found in a calculation with 48 spins (4x4 unit cells). This suggests, another state with an even larger unit cell may ultimately win the order by disorder competition. However, even if this is not the case, order by disorder due to fermion hopping and associated Berry flux has selected a 120$^0$ state that to our knowledge has never been considered before. Further it has more spins in its unit cell either the $\sqrt{3}\times\sqrt{3}$ state or the cuboc1 state. Hence, complex ground states can arise from an order by disorder mechanism.

There is also the question of whether there are other order by disorder mechanisms that generalize the case considered here and whether these would produce different decision making among the 120$^o$ manifold of states. The answer is likely yes: one could increase the spin representation of the fermion degree of freedom from spin $1/2$ to another spin $S$ as in the study of Ref. \onlinecite{Udagawa2013} who also consider the kagome lattice. This would enable the order by disorder effect to occur at other fillings than 1/2 with potentially different Berry flux desires. Spin representation could therefore introduce a hierarchy of order-by-disorder mechanisms each possibly selecting a different state. So, order by disorder via fermion hopping with Berry flux could provide a rich set of decision making capabilities on the kagome 120$^o$ manifold and other such manifolds common in highly frustrated magnetism.

%

\end{document}